\documentclass[twocolumn,times]{aastex63}
\usepackage[utf8]{inputenc}
\usepackage[normalem]{ulem}
\usepackage{amsmath}

\begin{document}

\title{Probing the Low-Velocity Regime of Non-Radiative Shocks with Neutron Star Bow Shocks}

\author[0000-0002-4941-5333]{Stella Koch Ocker}
\affiliation{Cahill Center for Astronomy and Astrophysics, California Institute of Technology, Pasadena, CA 91125, USA}
\affiliation{Carnegie Science Observatories, Pasadena, CA 91101, USA}
\author[0000-0002-2248-6107]{Maren Cosens}
\affiliation{Carnegie Science Observatories, Pasadena, CA 91101, USA}

\correspondingauthor{Stella Koch Ocker}
\email{socker@caltech.edu}

\begin{abstract}
   Non-radiative shocks accelerate particles and heat astrophysical plasmas. While supernova remnants are the most well-studied example, neutron star (NS) bow shocks are also non-radiative and Balmer-dominated. NS bow shocks are likely ubiquitous in the interstellar medium due to their large speeds imparted at birth, and they are thought to be a discrete source population contributing to the Galactic cosmic ray spectrum. To date, nine NS bow shocks have been directly observed in H$\alpha$ images. Most of these shocks have been characterized using narrowband H$\alpha$ imaging and slit spectroscopy, which do not resolve the multi-component velocity structure of the shocks and their spatial geometry. Here we present integral field spectroscopy of three NS bow shocks: J0742$-$2822, J1741$-$2054, and J2225$+$6535 (the Guitar Nebula). We measure the shock properties simultaneously in four dimensions: the 2D projected shock morphology, the radial velocity structure, and the H$\alpha$ flux. The broad-to-narrow line ratio ($I_{\rm b}/I_{\rm n}$) is inferred from radial velocity profiles, and for J1741$-$2054 the narrow line is detected in multiple regions of the shock. {The inferred line ratios and widths suggest that NS bow shocks represent a low shock velocity regime ($V \lesssim 200$ km/s) in which $I_{\rm b}/I_{\rm n}$ is high, distinct from the shock regime probed by supernova remnants. Our results illustrate a need for non-radiative shock models at velocities lower than previously considered, which will reveal the electron-ion temperature ratios and particle acceleration efficiencies of these bow shocks.} 
\end{abstract}


\section{Introduction}\label{sec:intro}

Optical imaging is the main method by which neutron star bow shocks are detected and characterized, as these shocks are non-radiative and chiefly emit in H$\alpha$ \citep{1988Natur.335..801K}. The H$\alpha$ emission arises from collisional excitation and charge exchange of the interstellar medium (ISM) where it confronts the pulsar wind. Only nine H$\alpha$-emitting neutron star bow shocks are known, likely because most of the neutron stars targeted in bow shock searches reside in regions of low neutral gas fraction \citep{2002ApJ...575..407C,brownsberger2014}, and because optical surveys are severely limited by extinction for neutron stars residing deep in the Galactic plane. Nevertheless, many neutron stars are expected to produce bow shocks because the majority have transverse velocities larger than the fast magnetosonic speed of the ISM \citep{verbunt2017}. 

Neutron star bow shocks are one of the few observed analogs of non-radiative, Balmer-dominated shocks in supernova remnants (SNRs; \citealt{heng2010}). Similar to SNRs, these shocks are expected to have a narrow H$\alpha$ line component (width $\sim 10$s km s$^{-1}$) at the radial velocity of the ambient ISM arising from collisional excitation of neutral hydrogen crossing the shock interface, and a broad H$\alpha$ line component (width $\sim 100$s km s$^{-1}$) arising from charge exchange between pre- and post-shock particles \citep{heng2010}. For bow shocks, these two emission components give rise to a triple-peaked radial velocity profile consisting of the narrow H$\alpha$ line flanked by the blue and red-shifted components of the broad line from the near and far sides of the shock \citep{romani2022}. The broad-to-narrow line intensity ratio ($I_{\rm b}/I_{\rm n}$) and the broad line width ($W_{\rm b}$) are sensitive diagnostics of the shock physics, including electron-ion temperature equilibration \citep{raymond2023} and particle energy distributions \citep{nikolic2013}. Moreover, $I_{\rm b}/I_{\rm n}$ and $W_{\rm b}$ constrain the physical velocity of the shock, which combined with proper motion measurements yield an independent distance estimate \citep{vanAdelsberg2008}. While $I_{\rm b}/I_{\rm n}$ and $W_{\rm b}$ are routinely constrained for SNRs (e.g. \citealt{raymond2023} and refs. therein), there is only one published measurement of $I_{\rm b}/I_{\rm n}$ for a neutron star bow shock to date \citep{romani2022}. 

Radial velocity profiles are also critical to resolving the physical geometry of the shock, as 2D images only constrain the shock in projection. The shock morphology is determined by conditions for ram pressure balance between the pulsar wind and the ISM \citep[e.g.][]{1996ApJ...459L..31W}. Velocity profiles are sensitive to the shock's inclination relative to the plane of the sky, which combined with a model for the shock structure yields the physical shock stand-off radius and the radial velocity of the neutron star, which cannot be inferred from pulsar timing or radio imaging \citep{romani2017,deVries2020,romani2022}. The ram pressure conditions required to explain observed neutron star bow shocks have been exploited to characterize the energetics of pulsar winds \citep{1988Natur.335..801K,1993Natur.362..133C,vanKerwijk2001} and their anisotropy \citep{romani2010}, pulsar distances and velocities \citep{jones2002,deVries2020}, the neutron star moment of inertia \citep{romani2017}, and ISM density fluctuations \citep{2004ApJ...600L..51C,ocker_bowshocks}. Most of these prior studies were based on narrowband H$\alpha$ imaging and slit spectroscopy, necessitating model-based assumptions of the shock inclination angle and subsequent constraints.

In this study, we leverage integral field spectroscopy (IFS) to spatially resolve the H$\alpha$ velocity structure of three neutron star bow shocks: J0742$-$2822, J1741$-$2054, and J2225$+$6535 (the Guitar Nebula). The observations are described in Section~\ref{sec:obs}. The radial velocity profiles are used to constrain $I_{\rm b}/I_{\rm n}$, $W_{\rm b}$, and the shock inclinations (Section~\ref{sec:results}). We interpret our results in the context of analogous shocks in SNRs {in Section~\ref{sec:discuss}.}

\section{Observations}\label{sec:obs}

We observed J0742$-$2822, J1741$-$2054, and J2225$+$6535 with the Keck Cosmic Web Imager (KCWI) on the Keck II Telescope on UT 2024-02-01 (J0742$-$2822) and UT 2024-07-12 (J1741$-$2054 and J2225$+$6535). All sources were observed with the RH1 grating and medium slicer, covering a field-of-view of $16\overset{\prime\prime}{.}5$ by $20\overset{\prime\prime}{.}4$ at resolutions of $0.25$ \AA\ and $0\overset{\prime\prime}{.}7$, although seeing was typically 1.5 to 3 times larger (the latter, worst seeing was on 2024-07-12). Nine 600 s exposures were obtained for J0742$-$2822 and J1741$-$2054 and eighteen 600 s exposures for J2225$+$6535. Offset sky fields were observed throughout each session in addition to flux standards GD50 and G138$-$31.

Data were reduced using the KCWI Data Reduction Pipeline (\url{https://pypi.org/project/kcwidrp/}; \citealt{kcwi_drp_ascl}), described in \cite{kcwi_drp}. Bias frames, internal flats, and dome flats were used to correct the science images, and cosmic ray rejection was performed using the Laplacian edge detection algorithm described in \cite{astroscrappy}. Continuum bars and arc lamps were used to perform the geometry and wavelength calibrations. A sky model was constructed from the offset fields and sky subtraction quality was verified using off-target regions. Flux standards GD50 and G138$-$31 were used to perform absolute flux calibration at an error level $<4\%$. Individual flux and wavelength calibrated exposures were combined using \texttt{drizzle} \citep{drizzle}.

\section{Results}\label{sec:results}

\begin{figure*}
    \centering
    \includegraphics[width=\textwidth]{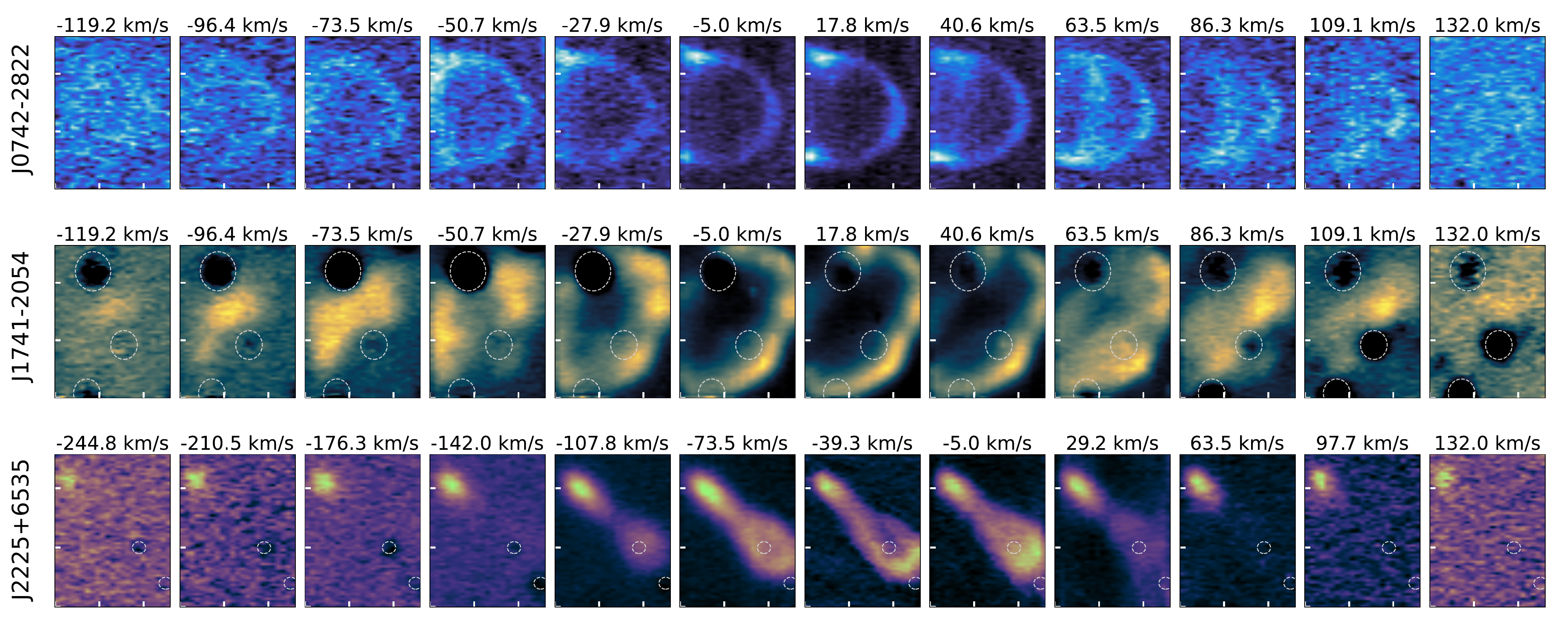}
    \caption{KCWI observations of three pulsar bow shocks, J0742$-$2822 (top), J1741$-$2054 (middle), and J2225$+$6535 (the Guitar Nebula; bottom). Images are shown for 12 radial velocity bins centered around the peak shock emission; the spectra have been smoothed from the native resolution of 0.25 \AA\ to show the large velocity range (larger bins are used for the Guitar due to its wider extension to negative velocities). These slices are continuum subtracted using off-band portions of the data cube; photospheric H$\alpha$ absorption causes many stars to show negative residuals, centered on different low velocity channels. Over-subtracted stars are shown by the grey dashed circles and are masked in subsequent analysis. The white tick marks are $5^{\prime\prime}$ apart; each image is about $13^{\prime\prime}$ on a side ($\approx$0.1 pc, 0.02 pc, and 0.04 pc on a side, from top to bottom). The dynamic range of each image is normalized to its peak flux level in order to illustrate shock structure at the lowest and highest velocities. }
    \label{fig:kcwi_panels}
\end{figure*}

\begin{figure*}
    \centering
    \includegraphics[width=\textwidth]{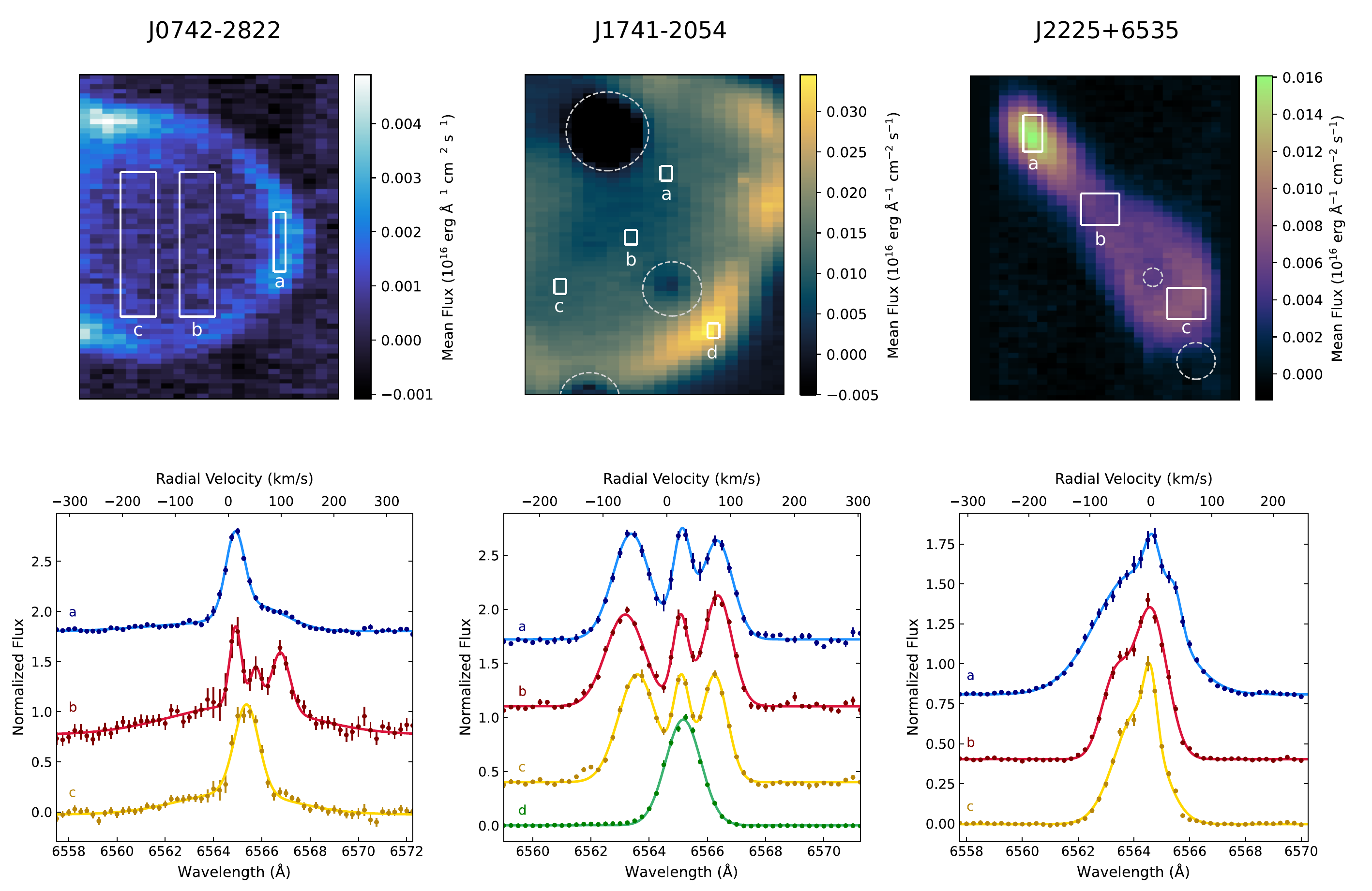}
    \caption{Heliocentric radial velocity profiles for J0742$-$2822 (left), J1741$-$2054 (middle), and J2225$+$6535 (right). The upper panels illustrate the regions of each shock used to calculate the aperture-averaged velocity profiles shown in the bottom panels, labeled (a,b,c,d). The images in the upper panels are averaged over the full velocity range of each shock, and the grey dashed circles indicate stars that are subsequently masked. The velocity profiles are each normalized to unity peak flux and shown with an arbitrary offset so that specific velocity features can be compared. Error bars correspond to the standard deviation of flux within each region at a given wavelength. Solid lines indicate the best-fit model for the velocity profile, which is treated as a sum of Gaussian components (\S~\ref{sec:results}).}
    \label{fig:velocity_profiles}
\end{figure*}

Figure~\ref{fig:kcwi_panels} shows our KCWI observations of the three bow shocks, illustrating the evolution of the shock emission over $\approx 6559 \rm \ \AA - 6568 \ \AA$, equivalent to radial velocities $\approx -250$ to $+150$ km/s (the exact range shown differs by shock). The spatial structure of the emission has a strong radial velocity dependence: limb brightening, lobes, and rings are apparent at different velocities and are indicative of changing conditions across the apex, body, and tail of each shock. We interpret these features using spectra averaged over different portions of the shocks. Figure~\ref{fig:velocity_profiles} shows the aperture-averaged velocity profiles and the regions of the shocks they correspond to. These regions were chosen to illustrate the highest signal-to-noise (S/N) narrow line detections and evolution of velocity features from shock apex to tail. The narrow line is detected in all three shocks. For J0742$-$2822 and J2225$+$6535, the narrow line is barely detected just behind the shock apexes, whereas for J1741$-$2054, the narrow line is detected across multiple parts of the shock. 

\subsection{Interpretation of Radial Velocity Profiles}

We model the velocity profiles as the sum of $N$ Gaussian components, where $N$ is determined by the number of distinct peaks in the velocity profile and validated through comparison of the reduced $\chi^2$ to that obtained for a model with one less component. The number of components required to explain each profile depends on spatial location, and in turn whether the faint narrow line is detected, which requires accurate sky subtraction. When the narrow line is detected, we estimate $I_{\rm b}/I_{\rm n}$ using the best-fit amplitudes of the narrow and broad components. We quote lower limits when the narrow line is covered by fewer than four resolution elements, i.e., it is not well separated from the broad line emission. The interpretation of $I_{\rm b}$ and $I_{\rm n}$ for bow shocks differs from the interpretation of shocks in SNRs. In the case of bow shocks, which are essentially conical shells, the velocity profile traces the line-of-sight through the shock viewed in projection. The observed narrow line intensity is thus the sum of the $I_{\rm n}$ contributions from both the near and far sides of the shock, both of which are at the same ambient ISM velocity. We therefore estimate $I_{\rm b}/I_{\rm n}$ by taking the sum of the blue and red-shifted broad line amplitudes ($I_{\rm b}^{\rm blue}$, $I_{\rm b}^{\rm red}$) and dividing by the amplitude of the narrow line:
\begin{equation}
    \frac{I_{\rm b}}{I_{\rm n}} \equiv \frac{I_{\rm b}^{\rm blue} + I_{\rm b}^{\rm red}}{I_{\rm n}}.
\end{equation}

Interpretation of the broad-line width $W_{\rm b}$ is less straightforward because the apparent line width is a function of inclination angle. When the shock is close to the plane of the sky ($i\approx0^\circ$), the widths of the blue and redshifted broad lines ($W_{\rm b}^{\rm blue}$, $W_{\rm b}^{\rm red}$) should be roughly equal. For inclination angles out of the plane of the sky, $W_{\rm b}^{\rm blue} > W_{\rm b}^{\rm red}$, while for inclination angles into the plane of the sky, $W_{\rm b}^{\rm blue} < W_{\rm b}^{\rm red}$. The exact dependence of $W_{\rm b}^{\rm blue}$ and $W_{\rm b}^{\rm red}$ on inclination will depend on the 3D spatial morphology of the shock, and hence will differ depending on the shock stand-off radius, the properties of the pulsar wind (e.g. whether it is equatorial or not), and the ISM density. Producing such 3D models for each shock is beyond the scope of this paper, but we nonetheless report empirical constraints on the apparent line width to quantitatively illustrate the velocity range of each shock. 

\begin{figure}
    \centering
    \includegraphics[width=0.45\textwidth]{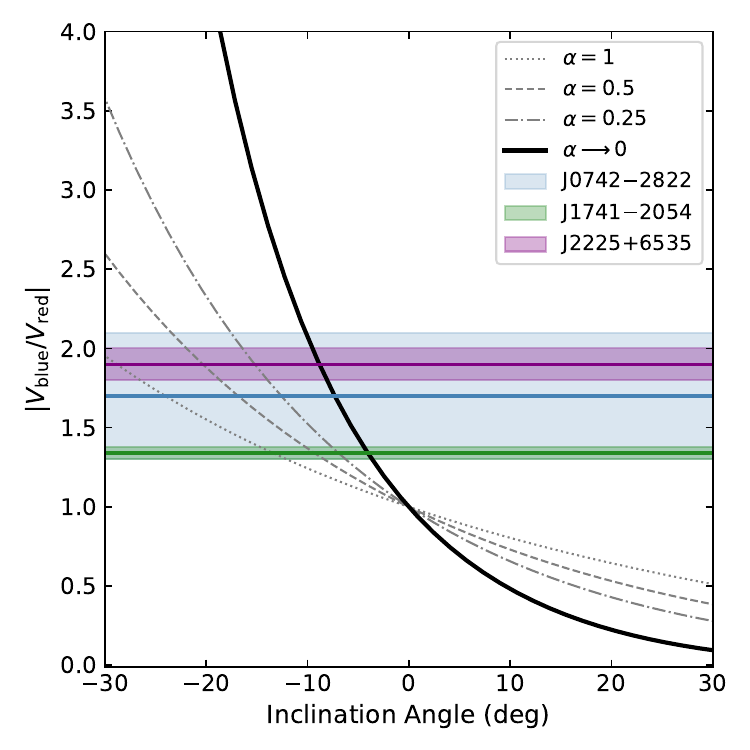}
    \caption{Ratio of blue to redshifted broad line radial velocity $|V_{\rm blue}/V_{\rm red}|$ vs. shock inclination angle. Negative angles correspond to inclination out of the sky and positive angles into the sky. The horizontal lines and shaded regions show inferred values and $1\sigma$ uncertainties of $|V_{\rm blue}/V_{\rm red}|$ for J0742$-$2822 (blue), J1741$-$2054 (green), and J2225$+$6535 (purple). The expected velocity ratio is estimated for an axisymmetric shock from an isotropic stellar wind \citep{1996ApJ...459L..31W}, for different values of $\alpha = V_*/V_w$, the ratio of the ISM velocity to stellar wind velocity. We constrain approximate ranges of inclination angle for each shock using the limit $\alpha \rightarrow 0$ (solid black curve), as appropriate for a relativistic wind.}
    \label{fig:inclination}
\end{figure}

Even in the absence of detailed shock models, we can place constraints on the plausible range of inclination angles based on the difference in velocity extrema of the blue and red-shifted broad lines ($V_{\rm blue}$, $V_{\rm red}$). If we assume an axisymmetric shock, then $|V_{\rm blue}| \approx |V_{\rm red}|$ when the inclination relative to the plane of the sky is $i\approx 0^\circ$. When $i < 0^\circ$ (out of the plane of the sky), $|V_{\rm blue}/V_{\rm red}| > 1$, whereas for $i > 0^\circ$ (into the plane of the sky), $|V_{\rm blue}/V_{\rm red}| < 1$. We use the thin-shell approximation for an isotropic stellar wind provided by \cite{1996ApJ...459L..31W} as a starting point, noting caveats below. In this scenario, the velocity tangent to the shock at the contact discontinuity is given by \cite{1996ApJ...459L..31W} as
\begin{equation}\label{eq:shock_velocity}
    v_t = v_* \frac{\sqrt{(\theta - \rm sin\theta cos\theta)^2 + (\tilde{w}^2 - sin^2\theta)^2)}}{2\alpha(1-\rm cos\theta) + \tilde{w}^2},
\end{equation}
where $\theta$ is the position angle along the shock measured relative to the shock apex, $v_*$ is the ISM velocity, $\alpha = v_*/v_w$ is the ratio of the ISM velocity to the stellar wind velocity ($\ll 1$ for relativistic pulsar winds), and $\rm \tilde{w}^2$ is given by
\begin{equation}
   \rm \tilde{w}^2 = 3(1 - \theta \rm cot\theta).
\end{equation}
We evaluate $v_t$ for a range of inclination angles and equivalent $\theta$, and estimate the resulting ratio of $V_{\rm blue}/V_{\rm red}$. Here we implicitly assume that the observed radial velocity depends on $\theta$ in the same fashion as $v_t$, normalized such that negative inclination $i$ corresponds to inclination out of the plane of the sky.
Note that the dependence of $v_t$ on $v_*$ does not matter in this context because we are only interested in the velocity ratio between the near and far sides of the shock. Figure~\ref{fig:inclination} shows $|V_{\rm blue}/V_{\rm red}|$ vs. $i$ for different values of $\alpha$; for relativistic pulsar winds, we use the limit $\alpha \rightarrow0$.  

In general, the velocity of the H$\alpha$ emission will not exactly follow that of the contact discontinuity; e.g., \cite{bucciantini1} find that while $v_t$ in the external layer where H$\alpha$ emission is produced (between the contact discontinuity and forward shock) tends to track the velocity of the contact discontinuity near the apex, $v_t$ in the external layer increases more rapidly in the tail of the shock. {The Wilkin model assumes conservation of momentum (isothermal shocks), whereas non-radiative shocks conserve energy and gas accelerates down the shock, leading to $v_t$ larger than in the isothermal case.}
Anisotropy in the pulsar wind and density variations in the ISM will further modify the velocity structure of the shock from the simple form in Equation~\ref{eq:shock_velocity} \citep{wilkin2000,2007MNRAS.374..793V}. The inclination angles given below should thus be regarded as fiducial estimates, to inform more detailed future modeling. 

\subsection{Individual Pulsar Results}

Table~\ref{tab:results} summarizes the quantitative constraints obtained from the radial velocity profiles. Results relevant to each pulsar are discussed below, along with information relevant to the interpretation of each measurement. 

\begin{deluxetable*}{c c c c c c | c c c c c}\label{tab:results}
    \tablecaption{Bow Shock Properties}
    \tabletypesize{\footnotesize}
    \tablehead{\colhead{PSR} & \colhead{Region} & \colhead{$I_{\rm b}/I_{\rm n}$} & \colhead{$V_{\rm n}$ (km/s)} & \colhead{$|V_{\rm blue}/V_{\rm red}|$} & \colhead{$W_{\rm b}$ (km/s)} & \colhead{$\mu$ (mas/yr)} & \colhead{Distance (kpc)} & \colhead{$V_{\perp}^{\rm psr}$ (km/s)} & \colhead{$\theta_{\rm BS}$ ($^{\prime\prime}$)} & \colhead{Refs.}}
    \startdata
    J0742$-$2822 & (b) & $\geq 3.0\pm0.5$ & $52 \pm 3$ & $1.7\pm0.4$ & $\lesssim160$ & $29\pm1$ & 2.07$^\dagger$ & 287 & 1.4 & 1, 2 \\ \hline
                 & (a) & $2.0\pm0.1$ & $23\pm1$ & $1.30\pm0.05$ & $62\pm6$ & & & & \\
    J1741$-$2054 & (b) & $2.3\pm0.2$ & $21\pm1$ & $1.39\pm0.02$ & $60\pm9$ & $109 \pm 10$ & 0.38$^\dagger$ & 197 & 2.3 & 1, 3 \\
                 & (c) & $2.2\pm0.1$ & $23\pm1$ & $1.33\pm0.03$ & $58 \pm 11$ & & & & \\ \hline 
    J2225$+$6535 & (a) & $\geq 3.1\pm0.1$ & $3.3\pm1.1$ & $1.9\pm0.1$ & $\lesssim 150$ & $194.1 \pm 0.3$ & 0.831 & 765 & 0.096 & 4, 5 \\
    \enddata
    \tablecomments{The broad-to-narrow line ratio $I_{\rm b}/I_{\rm n}$ is quoted as a lower limit when the narrow line S/N is low. The narrow line radial velocity $V_{\rm n}$ is reported in the heliocentric frame; for J0742$-$2822 and J1741$-$2054, the apparently substantial $V_{\rm n}$ are broadly consistent with Galactic differential rotation after correcting to the Local Standard of Rest. The broad-line velocity ratio $|V_{\rm blue}/V_{\rm red}|$ is obtained by calculating the velocity extrema of the blue and redshifted broad lines using the integrated flux of each broad line (\S~\ref{sec:results}). The broad line width $W_{\rm b}$ is reported as an upper limit when the blue and redshifted broad lines are highly asymmetric and not well separated; in all cases, $W_{\rm b}$ is based on the apparent full-width-at-half-maximum of the broad line and does not account for the effect of inclination. Quantities on the right half of the table are drawn from the literature. Distances indicated by $\dagger$ are estimated from NE2001 \citep{ne20011}; {pulsar transverse velocities ($V_{\perp}^{\rm psr}$) are estimated from the proper motion and distance}. The projected angular stand-off radii $\theta_{\rm BS}$ correspond to the most recently published values, and are based on differing shock models. References: (1) \cite{brownsberger2014}; (2) \cite{bailes1990}; (3) \cite{auchettl2015}; (4) \cite{deVries_Guitar}; (5) \cite{deller2019}.}
\end{deluxetable*}

\subsection{J0742$-$2822}

The shock emission of this pulsar is dominated by the blue-shifted broad line, which has a low-amplitude extension to large negative velocities, suggesting that the shock is viewed at substantial inclination out of the plane of the sky. A high inclination angle would also explain the ring feature that appears to move in velocity space to the right of the images shown in Figure~\ref{fig:kcwi_panels} \citep[e.g.][]{2020MNRAS.497.2605B}. The narrow line is weakly detected in Region (b). Due to its low S/N ($\approx 5$), we report a lower limit on the broad-to-narrow line ratio, $I_{\rm b}/I_{\rm n} \geq 3.0\pm0.5$. {Since we needed to average over a large region to detect the narrow line, the resulting spectrum covers a range of azimuthal angles, which broadens the apparent line width.}

Given the large asymmetry in the velocity profile, we place an upper limit on $W_{\rm b}$ using the full-width-at-half-maximum (FWHM) of the blueshifted broad line, $W_{\rm b} \lesssim 160$ km/s. To estimate the velocity ratio between the blue and redshifted broad lines, we first correct for the ambient ISM radial velocity measured from the narrow line. If we use the best-fit peaks of the broad lines, then we find $|V_{\rm blue}/V_{\rm red}| \approx 1$. However, this approach ignores the higher flux density of the blueshifted broad line and its extension to large negative velocities. We therefore calculate the integrated fluxes of the best-fit blue and red broad lines as functions of wavelength. To capture the velocity extrema, which are the most sensitive indicators of the shock inclination,  we then evaluate the wavelengths at which the integrated fluxes of the blue and redshifted lines reach $15\%$ and $85\%$ of their total values, respectively. Using this approach, which accounts for the actual partitioning of flux between the blue and redshifted lines, we find $V_{\rm blue} = -141 \pm 20$ km/s and $V_{\rm red} = 82 \pm 13$ km/s, yielding a velocity ratio $|V_{\rm blue}/V_{\rm red}| = 1.7\pm0.4$. For an axisymmetric shock and isotropic wind, this ratio suggests an inclination angle $-11^\circ \lesssim i \lesssim -2^\circ$ (see Figure~\ref{fig:inclination}). This inclination is consistent with the previously published upper limit of $|i| \lesssim 25^\circ$ \citep{jones2002}, which was based on the projected surface brightness required to see the shock in narrowband imaging.

\subsection{J1741$-$2054}

The overall morphology of this bow shock is highly complex. At large radial velocities, the shock emission stretches across two main lobes that are oriented perpendicular to the pulsar proper motion, which cuts diagonally from the upper left to bottom right of the images in Figure~\ref{fig:kcwi_panels} \citep{romani2010}. This two-lobed structure may support previous interpretations of narrowband H$\alpha$ images, which found that the flatness of the shock apex was consistent with an equatorial pulsar wind \citep{romani2010}.

The narrow line is detected in three regions behind the shock (a, b, and c). Using the best-fit amplitudes of the broad and narrow components, we find $I_{\rm b}/I_{\rm n} = 2.0 \pm 0.1$, $2.3\pm0.2$, and $2.2\pm0.1$ in Regions (a), (b) and (c), respectively. The radial velocity of the narrow line shifts by $<2$ km/s between each region (Table~\ref{tab:results}). Given these variations at the $1\sigma$ level, we find that $I_{\rm b}/I_{\rm n}$ is statistically consistent with being constant across the shock.

We constrain the broad line width using the FWHM averaged across both the blue and redshifted broad line components. We find $W_{\rm b}\approx 60$ km/s and is consistent with being constant across the three velocity profiles behind the shock apex (see Table~\ref{tab:results}). These widths are also consistent at the $1\sigma$ level with the FWHM of the single broad component detected at the shock apex in Region (d), which gives a width of $63.90\pm0.01$ km/s.

The high S/N and clear separation between the blue and red sides of the velocity profiles shown in Figure~\ref{fig:velocity_profiles} yield precise constraints on $|V_{\rm blue}/V_{\rm red}|$ shown in Table~\ref{tab:results}. For an axisymmetric shock, the mean value $|V_{\rm blue}/V_{\rm red}| = 1.34\pm0.04$ suggests an inclination angle $-6^\circ \lesssim i \lesssim -2^\circ$. An inclination angle out of the plane of the sky is supported by the blueshifted broad line width, which is up to 20 km/s wider than the redshifted broad line. This estimate is smaller but technically consistent with \cite{romani2010}, who found $i = -15\pm10^\circ$ based on the flatness of the shock apex and velocity shifts in split spectra.

\subsection{Guitar Nebula}

Our observation of the Guitar Nebula was substantially impacted by seeing, which blurs the shock's undulating morphology seen best in space-based images \citep{2002ApJ...575..407C,chatterjee2004,ocker_bowshocks,deVries_Guitar}. The narrow line is only detected in Region (a) close to the shock apex. Further down the shock body, the narrow line blends with the redshifted broad line, likely due to the shock inclination and the broad line width decreasing downstream from the shock apex. Given that seeing was substantially larger than the shock stand-off radius (Table~\ref{tab:results}), the narrow line may include flux from preshock ionization. We thus report a lower limit $I_{\rm b}/I_{\rm n} \geq 3.1 \pm 0.2$.

The radial velocity profile is asymmetrically skewed negative, again suggesting an inclination out of the plane of the sky. Given this large skew, we report an upper limit on the broad line width from Region (a), $W_{\rm b} \lesssim 150$ km/s, based on the FWHM of the blueshifted broad line. We constrain $|V_{\rm blue}/V_{\rm red}|$ in the same manner as for J0742$-$2822 and J1741$-$2054, finding $|V_{\rm blue}/V_{\rm red}| = 1.9\pm0.1$ and $-11^\circ \lesssim i \lesssim -7^\circ$. This inclination angle is larger than previous estimates because our data reveal a more significant blueshift than previous IFS observations \citep{deVries_Guitar}. Higher spatial-resolution IFS data (e.g. at similar resolution to \citealt{deVries_Guitar}) covering a larger portion of the shock could reveal concentric rings of H$\alpha$ emission where the shock is pinched, presumably due to ISM density fluctuations \citep{2020MNRAS.497.2605B}.

\section{Discussion}\label{sec:discuss}

The broad-line widths inferred above are all below the smallest values included in state-of-the-art simulations of non-radiative, Balmer-dominated shocks \citep[e.g.][]{ghavamian2001,vanAdelsberg2008}, implying that the bow shocks fall into a low shock velocity regime. This conclusion is further compounded by the inferred inclination angles, which suggest that the widths reported in Table~\ref{tab:results} are overestimates of the true, de-projected $W_{\rm b}$. 

\begin{figure}
    \centering
    \includegraphics[width=0.45\textwidth]{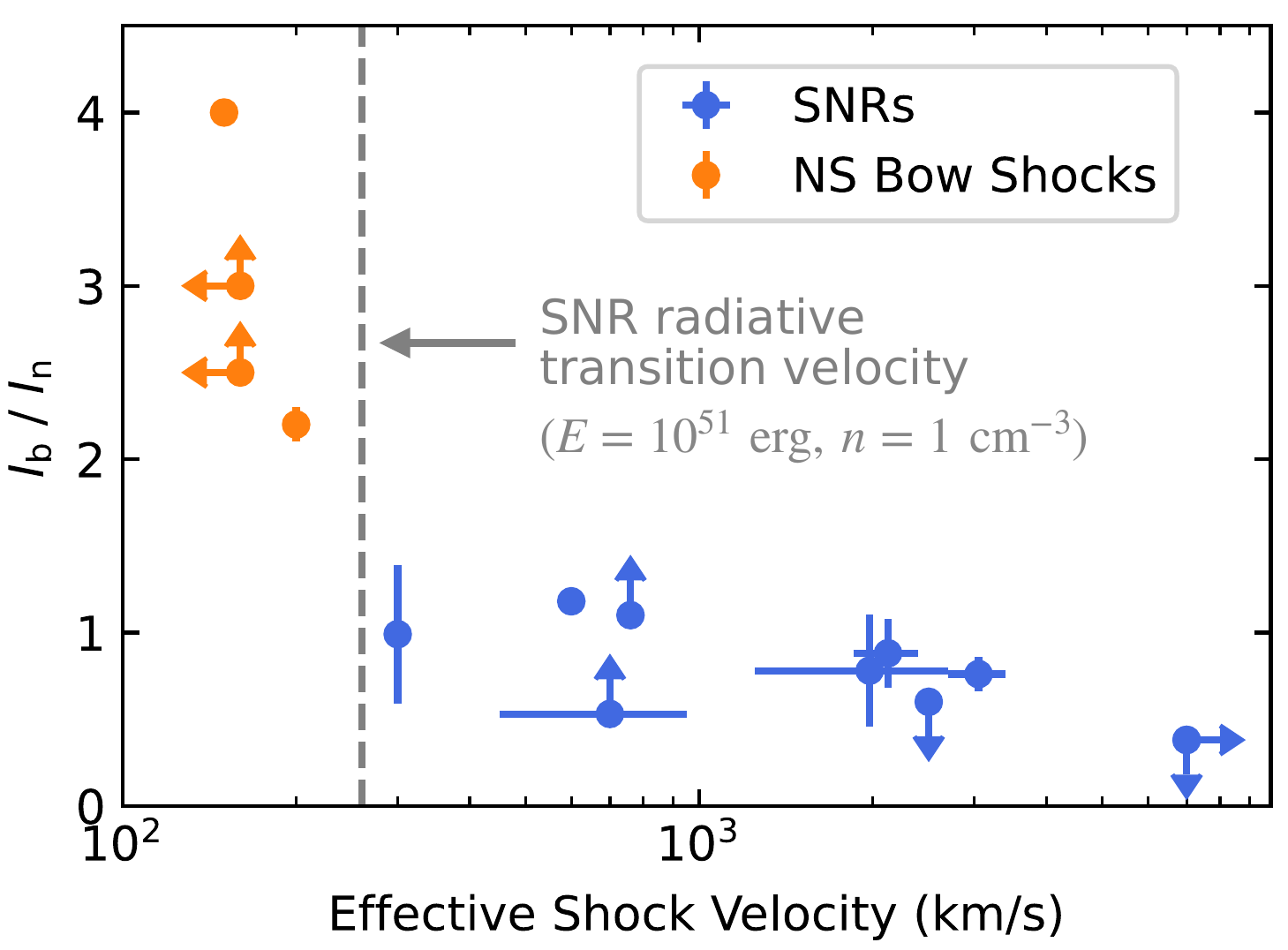}
    \caption{{Broad-to-narrow line ratio ($I_{\rm b}/I_{\rm n}$) vs. effective shock velocity for both SNRs (blue) and neutron star bow shocks (orange). SNR constraints are from the recent literature compilation of \cite{raymond2023}, and orange points indicate values inferred for the three neutron star bow shocks in this study and one bow shock from \cite{romani2022}. All four bow shocks appear to have velocities below the SNR radiative transition, which is shown by the grey dashed line for a nominal energy $E = 10^{51}$ erg and density $n = 1$ cm$^{-3}$ \citep{blondin98}.}}
    \label{fig:beta}
\end{figure}

Following \cite{romani2022}, we use the relationship between H$\alpha$ production efficiencies and $I_{\rm b}/I_{\rm n}$ modeled by \cite{hengmccray2007} to estimate the effective shock velocity $v_s$. {Figure~\ref{fig:beta} shows $I_{\rm b}/I_{\rm n}$ vs. $v_s$ for the three neutron star bow shocks observed with KCWI and the one other neutron star bow shock with a detection of both broad and narrow lines, J1959$+$2048 \citep{romani2022}.} For J1741$-$2054, $I_{\rm b}/I_{\rm n} \approx 2$, and we find $v_s\approx 200$ km/s. The pulsar proper motion and estimated distance give a transverse velocity of $197$ km/s, which is broadly consistent with the effective shock velocity. {The limits on $I_{\rm b}/I_{\rm n}$ for J0742$-$2822 and the Guitar Nebula similarly yield low velocities ($v_s \lesssim 160$ km/s). Published non-radiative models do not treat this low-velocity regime because SNR shocks become radiative at a transition velocity $V_{\rm trans} \approx 260\ E_{51}^{1/17}n^{2/17}\ {\rm km/s}$, where $E_{51}$ is energy in units of $10^{51}$ ergs and $n$ is density in units of cm$^{-3}$ \citep{blondin98}. Below this transition velocity, hydrogen is ionized before it reaches the SNR shock, and the broad-to-narrow H$\alpha$ line diagnostic cannot be measured. The closest available shock model is \cite{ghavamian2001}, which models the dependence of of $I_{\rm b}/I_{\rm n}$ on electron-ion equilibration for a lowest shock velocity $v_s \approx 250$ km/s. Naively applying this model to the $I_{\rm b}/I_{\rm n}$ inferred here would yield electron-ion temperature ratios $\beta = T_e/T_p \lesssim 0.05$, dramatically smaller than values expected for SNRs, which follow a well-established trend in which $\beta$ declines with shock velocity \citep{ghavamian2013, raymond2023}. This discrepancy may be reconciled in part if bow shocks correspond to lower temperatures than SNRs, because $I_{\rm b}/I_{\rm n}$ depends on the relative rates of charge transfer (which varies slowly with shock speed) and collisional excitation (which will be smaller for lower temperatures). Additionally, \cite{ghavamian2001} assume that no Ly$\beta$ is converted to H$\alpha$, which may not be true for small broad line widths. Allowing for Ly$\beta$ conversion could further increase $I_{\rm b}/I_{\rm n}$ from the \cite{ghavamian2001} predictions. Accounting for all of these effects, in addition to the potential role of Coulomb collisions, will be critical to accurately determine $\beta$ for the bow shock properties we infer.}

{Our results illustrate a strong need for both non-radiative shock simulations at low velocities ($\lesssim 200$ km/s) and high-resolution IFS observations of the full neutron star bow shock sample, in order to establish whether these sources represent a distinct regime of $\beta$ and $v_s$ from SNRs.} The electron-ion temperature ratio is a fundamental diagnostic of nonthermal electron acceleration, and of the degree to which plasma turbulence transfers energy between ions and electrons. In SNRs, the decrease in $\beta$ with higher $v_s$ has been interpreted in the context of diffusive shock acceleration as a decline in electron injection efficiency with increasing shock speed \citep{morlino2021}, which can ultimately contribute to the differing slopes of observed cosmic ray electron and proton spectra \citep{aguilar2019a,aguilar2019b}. 
Pulsar wind nebulae (PWNe) play a critical role in cosmic ray acceleration across a broad energy range (10s Gev - TeV). They are commonly invoked to explain the local positron excess \citep{abeysekara2017,petrov2020,dimauro2021,orusa2021}. The role of discrete sources in observed cosmic ray spectra continues to be a topic of debate \citep[e.g.][]{butsky2024}, with neutron stars serving as one of the main source populations of interest \citep[e.g.][]{thaler2023}. PWNe with bow shocks represent an important subset of neutron stars' contribution to the cosmic ray spectrum, as reacceleration of ultrarelativistic particles at the bow shock interface can contribute a significant fraction of positrons detected at $\gtrsim 10$s GeV \citep{bykov2017}. While the escape of these particles can be directly traced through UV and X-ray observations \citep{2017ApJ...835..264R,deVries2022_J2030}, the H$\alpha$ velocity structure offers a critical additional handle on the underlying physics, not only by resolving the 3D geometry of the wind and shock, but also by constraining the conditions for electron-proton equilibration and, with sufficient resolution, the underlying particle distributions. {Further expanding IFS observations to the tails of these shocks may additionally constrain the role of mass-loading in shaping these shocks' larger-scale, undulating morphologies, the source of which remains debated \citep{morlino2015}.}

\facility{Keck:II (KCWI)}

\software{\texttt{KCWI\_DRP} \citep{kcwi_drp_ascl}, \texttt{drizzle} \citep{drizzle}}

\acknowledgements{The authors thank John Raymond, Roger Romani, and Leon Garcia for their invaluable input on this work. SKO and MC are supported by the Brinson Foundation through the Brinson Prize Fellowship Program. SKO is a member of the NANOGrav Physics Frontiers Center (NSF award PHY-2020265). Part of this work was performed at the Aspen Center for Physics, which is supprted by National Science Foundation grant PHY-2210452. Some of the data presented herein were obtained at Keck Observatory (Keck Program C422, PI: Ocker), which is a private 501(c)3 non-profit organization operated as a scientific partnership among the California Institute of Technology, the University of California, and the National Aeronautics and Space Administration. The Observatory was made possible by the generous financial support of the W. M. Keck Foundation. Rosalie McGurk and Michael McDonald gave advice on observing and data reduction. Caltech and Carnegie Observatories are located on the traditional and unceded lands of the Tongva people. The authors wish to recognize and acknowledge the very significant cultural role and reverence that the summit of Maunakea has always had within the Native Hawaiian community. We are most fortunate to have the opportunity to conduct observations from this mountain.}

\bibliography{master_bib}

\begin{thebibliography}{}
\expandafter\ifx\csname natexlab\endcsname\relax\def\natexlab#1{#1}\fi
\providecommand{\url}[1]{\href{#1}{#1}}
\providecommand{\dodoi}[1]{doi:~\href{http://doi.org/#1}{\nolinkurl{#1}}}
\providecommand{\doeprint}[1]{\href{http://ascl.net/#1}{\nolinkurl{http://ascl.net/#1}}}
\providecommand{\doarXiv}[1]{\href{https://arxiv.org/abs/#1}{\nolinkurl{https://arxiv.org/abs/#1}}}

\bibitem[{{Abeysekara} {et~al.}(2017){Abeysekara}, {Albert}, {Alfaro}, {Alvarez}, {{\'A}lvarez}, {Arceo}, {Arteaga-Vel{\'a}zquez}, {Avila Rojas}, {Ayala Solares}, {Barber}, {Bautista-Elivar}, {Becerril}, {Belmont-Moreno}, {BenZvi}, {Berley}, {Bernal}, {Braun}, {Brisbois}, {Caballero-Mora}, {Capistr{\'a}n}, {Carrami{\~n}ana}, {Casanova}, {Castillo}, {Cotti}, {Cotzomi}, {Couti{\~n}o de Le{\'o}n}, {De Le{\'o}n}, {De la Fuente}, {Dingus}, {DuVernois}, {D{\'\i}az-V{\'e}lez}, {Ellsworth}, {Engel}, {Enr{\'\i}quez-Rivera}, {Fiorino}, {Fraija}, {Garc{\'\i}a-Gonz{\'a}lez}, {Garfias}, {Gerhardt}, {Gonz{\'a}lez Mu{\~n}oz}, {Gonz{\'a}lez}, {Goodman}, {Hampel-Arias}, {Harding}, {Hern{\'a}ndez}, {Hern{\'a}ndez-Almada}, {Hinton}, {Hona}, {Hui}, {H{\"u}ntemeyer}, {Iriarte}, {Jardin-Blicq}, {Joshi}, {Kaufmann}, {Kieda}, {Lara}, {Lauer}, {Lee}, {Lennarz}, {Vargas}, {Linnemann}, {Longinotti}, {Luis Raya}, {Luna-Garc{\'\i}a}, {L{\'o}pez-Coto}, {Malone}, {Marinelli}, {Martinez}, {Martinez-Castellanos}, {Mart{\'\i}nez-Castro},
  {Mart{\'\i}nez-Huerta}, {Matthews}, {Miranda-Romagnoli}, {Moreno}, {Mostaf{\'a}}, {Nellen}, {Newbold}, {Nisa}, {Noriega-Papaqui}, {Pelayo}, {Pretz}, {P{\'e}rez-P{\'e}rez}, {Ren}, {Rho}, {Rivi{\`e}re}, {Rosa-Gonz{\'a}lez}, {Rosenberg}, {Ruiz-Velasco}, {Salazar}, {Salesa Greus}, {Sandoval}, {Schneider}, {Schoorlemmer}, {Sinnis}, {Smith}, {Springer}, {Surajbali}, {Taboada}, {Tibolla}, {Tollefson}, {Torres}, {Ukwatta}, {Vianello}, {Weisgarber}, {Westerhoff}, {Wisher}, {Wood}, {Yapici}, {Yodh}, {Younk}, {Zepeda}, {Zhou}, {Guo}, {Hahn}, {Li}, \& {Zhang}}]{abeysekara2017}
{Abeysekara}, A.~U., {Albert}, A., {Alfaro}, R., {et~al.} 2017, Science, 358, 911, \dodoi{10.1126/science.aan4880}

\bibitem[{{Aguilar} {et~al.}(2019{\natexlab{a}}){Aguilar}, {Ali Cavasonza}, {Ambrosi}, {Arruda}, {Attig}, {Azzarello}, {Bachlechner}, {Barao}, {Barrau}, {Barrin}, {Bartoloni}, {Basara}, {Ba{\c{s}}e{\v{g}}mez-du Pree}, {Battiston}, {Becker}, {Behlmann}, {Beischer}, {Berdugo}, {Bertucci}, {Bindi}, {de Boer}, {Bollweg}, {Borgia}, {Boschini}, {Bourquin}, {Bueno}, {Burger}, {Burger}, {Cai}, {Capell}, {Caroff}, {Casaus}, {Castellini}, {Cervelli}, {Chang}, {Chen}, {Chen}, {Chen}, {Cheng}, {Chou}, {Choutko}, {Chung}, {Clark}, {Coignet}, {Consolandi}, {Contin}, {Corti}, {Crispoltoni}, {Cui}, {Dadzie}, {Dai}, {Datta}, {Delgado}, {Della Torre}, {Demirk{\"o}z}, {Derome}, {Di Falco}, {Dimiccoli}, {D{\'\i}az}, {von Doetinchem}, {Dong}, {Donnini}, {Duranti}, {Egorov}, {Eline}, {Eronen}, {Feng}, {Fiandrini}, {Fisher}, {Formato}, {Galaktionov}, {Garc{\'\i}a-L{\'o}pez}, {Gargiulo}, {Gast}, {Gebauer}, {Gervasi}, {Giovacchini}, {G{\'o}mez-Coral}, {Gong}, {Goy}, {Grabski}, {Grandi}, {Graziani}, {Guo}, {Haino}, {Han}, {He},
  {Heil}, {Hsieh}, {Huang}, {Huang}, {Incagli}, {Jia}, {Jinchi}, {Kanishev}, {Khiali}, {Kirn}, {Konak}, {Kounina}, {Kounine}, {Koutsenko}, {Kulemzin}, {La Vacca}, {Laudi}, {Laurenti}, {Lazzizzera}, {Lebedev}, {Lee}, {Lee}, {Leluc}, {Li}, {Li}, {Li}, {Li}, {Light}, {Lin}, {Lippert}, {Liu}, {Liu}, {Liu}, {Lu}, {Lu}, {Luebelsmeyer}, {Luo}, {Luo}, {Luo}, {Lyu}, {Machate}, {Ma{\~n}{\'a}}, {Mar{\'\i}n}, {Martin}, {Mart{\'\i}nez}, {Masi}, {Maurin}, {Menchaca-Rocha}, {Meng}, {Mo}, {Molero}, {Mott}, {Mussolin}, {Nelson}, {Ni}, {Nikonov}, {Nozzoli}, {Oliva}, {Orcinha}, {Palermo}, {Palmonari}, {Paniccia}, {Pashnin}, {Pauluzzi}, {Pensotti}, {Perrina}, {Phan}, {Picot-Clemente}, {Plyaskin}, {Pohl}, {Poireau}, {Popkow}, {Quadrani}, {Qi}, {Qin}, {Qu}, {Rancoita}, {Rapin}, {Conde}, {Rosier-Lees}, {Rozhkov}, {Rozza}, {Sagdeev}, {Solano}, {Schael}, {Schmidt}, {Schulz von Dratzig}, {Schwering}, {Seo}, {Shan}, {Shi}, {Siedenburg}, {Song}, {Sun}, {Tacconi}, {Tang}, {Tang}, {Tian}, {Ting}, {Ting}, {Tomassetti}, {Torsti}, {Urban},
  {Vagelli}, {Valente}, {Valtonen}, {V{\'a}zquez Acosta}, {Vecchi}, {Velasco}, {Vialle}, {Viz{\'a}n}, {Wang}, {Wang}, {Wang}, {Wang}, {Wang}, {Wang}, {Wei}, {Weng}, {Wu}, {Xiong}, {Xu}, {Yan}, {Yang}, {Yi}, {Yu}, {Yu}, {Zannoni}, {Zeissler}, {Zhang}, {Zhang}, {Zhang}, {Zhang}, {Zhao}, {Zheng}, {Zhuang}, {Zhukov}, {Zichichi}, {Zimmermann}, {Zuccon}, \& {AMS Collaboration}}]{aguilar2019a}
{Aguilar}, M., {Ali Cavasonza}, L., {Ambrosi}, G., {et~al.} 2019{\natexlab{a}}, PRL, 122, 041102, \dodoi{10.1103/PhysRevLett.122.041102}

\bibitem[{{Aguilar} {et~al.}(2019{\natexlab{b}}){Aguilar}, {Ali Cavasonza}, {Alpat}, {Ambrosi}, {Arruda}, {Attig}, {Azzarello}, {Bachlechner}, {Barao}, {Barrau}, {Barrin}, {Bartoloni}, {Basara}, {Ba{\c{s}}e{\v{g}}mez-du Pree}, {Battiston}, {Becker}, {Behlmann}, {Beischer}, {Berdugo}, {Bertucci}, {Bindi}, {de Boer}, {Bollweg}, {Borgia}, {Boschini}, {Bourquin}, {Bueno}, {Burger}, {Burger}, {Cai}, {Capell}, {Caroff}, {Casaus}, {Castellini}, {Cervelli}, {Chang}, {Chen}, {Chen}, {Chen}, {Cheng}, {Chou}, {Choutko}, {Chung}, {Clark}, {Coignet}, {Consolandi}, {Contin}, {Corti}, {Crispoltoni}, {Cui}, {Dadzie}, {Dai}, {Datta}, {Delgado}, {Della Torre}, {Demirk{\"o}z}, {Derome}, {Di Falco}, {Di Felice}, {Dimiccoli}, {D{\'\i}az}, {von Doetinchem}, {Dong}, {Donnini}, {Duranti}, {Egorov}, {Eline}, {Eronen}, {Feng}, {Fiandrini}, {Fisher}, {Formato}, {Galaktionov}, {Garc{\'\i}a-L{\'o}pez}, {Gargiulo}, {Gast}, {Gebauer}, {Gervasi}, {Giovacchini}, {G{\'o}mez-Coral}, {Gong}, {Goy}, {Grabski}, {Grandi}, {Graziani}, {Guo},
  {Haino}, {Han}, {He}, {Heil}, {Hsieh}, {Huang}, {Huang}, {Incagli}, {Jia}, {Jinchi}, {Kanishev}, {Khiali}, {Kirn}, {Konak}, {Kounina}, {Kounine}, {Koutsenko}, {Kulemzin}, {La Vacca}, {Laudi}, {Laurenti}, {Lazzizzera}, {Lebedev}, {Lee}, {Lee}, {Leluc}, {Li}, {Li}, {Li}, {Li}, {Light}, {Lin}, {Lippert}, {Liu}, {Liu}, {Liu}, {Lu}, {Lu}, {Luebelsmeyer}, {Luo}, {Luo}, {Luo}, {Lyu}, {Machate}, {Ma{\~n}{\'a}}, {Mar{\'\i}n}, {Martin}, {Mart{\'\i}nez}, {Masi}, {Maurin}, {Menchaca-Rocha}, {Meng}, {Mo}, {Molero}, {Mott}, {Mussolin}, {Nelson}, {Ni}, {Nikonov}, {Nozzoli}, {Oliva}, {Orcinha}, {Palermo}, {Palmonari}, {Paniccia}, {Pashnin}, {Pauluzzi}, {Pensotti}, {Perrina}, {Phan}, {Picot-Clemente}, {Plyaskin}, {Pohl}, {Poireau}, {Popkow}, {Quadrani}, {Qi}, {Qin}, {Qu}, {Rancoita}, {Rapin}, {Conde}, {Rosier-Lees}, {Rozhkov}, {Rozza}, {Sagdeev}, {Solano}, {Schael}, {Schmidt}, {von Dratzig}, {Schwering}, {Seo}, {Shan}, {Shi}, {Siedenburg}, {Song}, {Sun}, {Tacconi}, {Tang}, {Tang}, {Tian}, {Ting}, {Ting}, {Tomassetti},
  {Torsti}, {Urban}, {Vagelli}, {Valente}, {Valtonen}, {Acosta}, {Vecchi}, {Velasco}, {Vialle}, {Viz{\'a}n}, {Wang}, {Wang}, {Wang}, {Wang}, {Wang}, {Wang}, {Wei}, {Weng}, {Wu}, {Xiong}, {Xu}, {Yan}, {Yang}, {Yi}, {Yu}, {Yu}, {Zannoni}, {Zeissler}, {Zhang}, {Zhang}, {Zhang}, {Zhang}, {Zhao}, {Zheng}, {Zhuang}, {Zhukov}, {Zichichi}, {Zimmermann}, {Zuccon}, \& {AMS Collaboration}}]{aguilar2019b}
{Aguilar}, M., {Ali Cavasonza}, L., {Alpat}, B., {et~al.} 2019{\natexlab{b}}, PRL, 122, 101101, \dodoi{10.1103/PhysRevLett.122.101101}

\bibitem[{{Auchettl} {et~al.}(2015){Auchettl}, {Slane}, {Romani}, {Posselt}, {Pavlov}, {Kargaltsev}, {Ng}, {Temim}, {Weisskopf}, {Bykov}, \& {Swartz}}]{auchettl2015}
{Auchettl}, K., {Slane}, P., {Romani}, R.~W., {et~al.} 2015, ApJ, 802, 68, \dodoi{10.1088/0004-637X/802/1/68}

\bibitem[{{Bailes} {et~al.}(1990){Bailes}, {Manchester}, {Kesteven}, {Norris}, \& {Reynolds}}]{bailes1990}
{Bailes}, M., {Manchester}, R.~N., {Kesteven}, M.~J., {Norris}, R.~P., \& {Reynolds}, J.~E. 1990, MNRAS, 247, 322

\bibitem[{{Barkov} {et~al.}(2020){Barkov}, {Lyutikov}, \& {Khangulyan}}]{2020MNRAS.497.2605B}
{Barkov}, M.~V., {Lyutikov}, M., \& {Khangulyan}, D. 2020, MNRAS, 497, 2605, \dodoi{10.1093/mnras/staa1601}

\bibitem[{{Blondin} {et~al.}(1998){Blondin}, {Wright}, {Borkowski}, \& {Reynolds}}]{blondin98}
{Blondin}, J.~M., {Wright}, E.~B., {Borkowski}, K.~J., \& {Reynolds}, S.~P. 1998, ApJ, 500, 342, \dodoi{10.1086/305708}

\bibitem[{{Brownsberger} \& {Romani}(2014)}]{brownsberger2014}
{Brownsberger}, S., \& {Romani}, R.~W. 2014, ApJ, 784, 154, \dodoi{10.1088/0004-637X/784/2/154}

\bibitem[{{Bucciantini}(2002)}]{bucciantini1}
{Bucciantini}, N. 2002, A\&A, 387, 1066, \dodoi{10.1051/0004-6361:20020495}

\bibitem[{{Butsky} {et~al.}(2024){Butsky}, {Hopkins}, {Kempski}, {Ponnada}, {Quataert}, \& {Squire}}]{butsky2024}
{Butsky}, I.~S., {Hopkins}, P.~F., {Kempski}, P., {et~al.} 2024, MNRAS, 528, 4245, \dodoi{10.1093/mnras/stae276}

\bibitem[{{Bykov} {et~al.}(2017){Bykov}, {Amato}, {Petrov}, {Krassilchtchikov}, \& {Levenfish}}]{bykov2017}
{Bykov}, A.~M., {Amato}, E., {Petrov}, A.~E., {Krassilchtchikov}, A.~M., \& {Levenfish}, K.~P. 2017, SSR, 207, 235, \dodoi{10.1007/s11214-017-0371-7}

\bibitem[{{Chatterjee} \& {Cordes}(2002)}]{2002ApJ...575..407C}
{Chatterjee}, S., \& {Cordes}, J.~M. 2002, ApJ, 575, 407, \dodoi{10.1086/341139}

\bibitem[{{Chatterjee} \& {Cordes}(2004)}]{2004ApJ...600L..51C}
---. 2004, ApJL, 600, L51, \dodoi{10.1086/381498}

\bibitem[{{Chatterjee} {et~al.}(2004){Chatterjee}, {Cordes}, {Vlemmings}, {Arzoumanian}, {Goss}, \& {Lazio}}]{chatterjee2004}
{Chatterjee}, S., {Cordes}, J.~M., {Vlemmings}, W.~H.~T., {et~al.} 2004, ApJ, 604, 339, \dodoi{10.1086/381748}

\bibitem[{{Cordes} \& {Lazio}(2002)}]{ne20011}
{Cordes}, J.~M., \& {Lazio}, T.~J.~W. 2002, arXiv e-prints, astro.
\newblock \doarXiv{astro-ph/0207156}

\bibitem[{{Cordes} {et~al.}(1993){Cordes}, {Romani}, \& {Lundgren}}]{1993Natur.362..133C}
{Cordes}, J.~M., {Romani}, R.~W., \& {Lundgren}, S.~C. 1993, Nature, 362, 133, \dodoi{10.1038/362133a0}

\bibitem[{{de Vries} \& {Romani}(2020)}]{deVries2020}
{de Vries}, M., \& {Romani}, R.~W. 2020, ApJL, 896, L7, \dodoi{10.3847/2041-8213/ab9640}

\bibitem[{{de Vries} \& {Romani}(2022)}]{deVries2022_J2030}
---. 2022, ApJ, 928, 39, \dodoi{10.3847/1538-4357/ac5739}

\bibitem[{{de Vries} {et~al.}(2022){de Vries}, {Romani}, {Kargaltsev}, {Pavlov}, {Posselt}, {Slane}, {Bucciantini}, {Ng}, \& {Klingler}}]{deVries_Guitar}
{de Vries}, M., {Romani}, R.~W., {Kargaltsev}, O., {et~al.} 2022, ApJ, 939, 70, \dodoi{10.3847/1538-4357/ac9794}

\bibitem[{{Deller} {et~al.}(2019){Deller}, {Goss}, {Brisken}, {Chatterjee}, {Cordes}, {Janssen}, {Kovalev}, {Lazio}, {Petrov}, {Stappers}, \& {Lyne}}]{deller2019}
{Deller}, A.~T., {Goss}, W.~M., {Brisken}, W.~F., {et~al.} 2019, ApJ, 875, 100, \dodoi{10.3847/1538-4357/ab11c7}

\bibitem[{{Di Mauro} {et~al.}(2021){Di Mauro}, {Donato}, \& {Manconi}}]{dimauro2021}
{Di Mauro}, M., {Donato}, F., \& {Manconi}, S. 2021, PRD, 104, 083012, \dodoi{10.1103/PhysRevD.104.083012}

\bibitem[{{Ghavamian} {et~al.}(2001){Ghavamian}, {Raymond}, {Smith}, \& {Hartigan}}]{ghavamian2001}
{Ghavamian}, P., {Raymond}, J., {Smith}, R.~C., \& {Hartigan}, P. 2001, ApJ, 547, 995, \dodoi{10.1086/318408}

\bibitem[{{Ghavamian} {et~al.}(2013){Ghavamian}, {Schwartz}, {Mitchell}, {Masters}, \& {Laming}}]{ghavamian2013}
{Ghavamian}, P., {Schwartz}, S.~J., {Mitchell}, J., {Masters}, A., \& {Laming}, J.~M. 2013, SSR, 178, 633, \dodoi{10.1007/s11214-013-9999-0}

\bibitem[{{Gonzaga} {et~al.}(2012){Gonzaga}, {Hack}, {Fruchter}, \& {Mack}}]{drizzle}
{Gonzaga}, S., {Hack}, W., {Fruchter}, A., \& {Mack}, J. 2012, {The DrizzlePac Handbook} (STSci)

\bibitem[{{Heng}(2010)}]{heng2010}
{Heng}, K. 2010, PASA, 27, 23, \dodoi{10.1071/AS09057}

\bibitem[{{Heng} \& {McCray}(2007)}]{hengmccray2007}
{Heng}, K., \& {McCray}, R. 2007, ApJ, 654, 923, \dodoi{10.1086/509601}

\bibitem[{{Jones} {et~al.}(2002){Jones}, {Stappers}, \& {Gaensler}}]{jones2002}
{Jones}, D.~H., {Stappers}, B.~W., \& {Gaensler}, B.~M. 2002, A\&A, 389, L1, \dodoi{10.1051/0004-6361:20020651}

\bibitem[{{Kulkarni} \& {Hester}(1988)}]{1988Natur.335..801K}
{Kulkarni}, S.~R., \& {Hester}, J.~J. 1988, Nature, 335, 801, \dodoi{10.1038/335801a0}

\bibitem[{{Morlino} \& {Celli}(2021)}]{morlino2021}
{Morlino}, G., \& {Celli}, S. 2021, MNRAS, 508, 6142, \dodoi{10.1093/mnras/stab2972}

\bibitem[{{Morlino} {et~al.}(2015){Morlino}, {Lyutikov}, \& {Vorster}}]{morlino2015}
{Morlino}, G., {Lyutikov}, M., \& {Vorster}, M. 2015, MNRAS, 454, 3886, \dodoi{10.1093/mnras/stv2189}

\bibitem[{{Morrissey} {et~al.}(2018){Morrissey}, {Matuszewski}, {Martin}, {Neill}, {Epps}, {Fucik}, {Weber}, {Darvish}, {Adkins}, {Allen}, {Bartos}, {Belicki}, {Cabak}, {Callahan}, {Cowley}, {Crabill}, {Deich}, {Delecroix}, {Doppman}, {Hilyard}, {James}, {Kaye}, {Kokorowski}, {Kwok}, {Lanclos}, {Milner}, {Moore}, {O'Sullivan}, {Parihar}, {Park}, {Phillips}, {Rizzi}, {Rockosi}, {Rodriguez}, {Salaun}, {Seaman}, {Sheikh}, {Weiss}, \& {Zarzaca}}]{kcwi_drp}
{Morrissey}, P., {Matuszewski}, M., {Martin}, D.~C., {et~al.} 2018, ApJ, 864, 93, \dodoi{10.3847/1538-4357/aad597}

\bibitem[{{Neill} {et~al.}(2023){Neill}, {Matuszewski}, {Martin}, {Brodheim}, \& {Rizzi}}]{kcwi_drp_ascl}
{Neill}, D., {Matuszewski}, M., {Martin}, C., {Brodheim}, M., \& {Rizzi}, L. 2023, {KCWI\_DRP: Keck Cosmic Web Imager Data Reduction Pipeline in Python}, Astrophysics Source Code Library, record ascl:2301.019

\bibitem[{{Nikoli{\'c}} {et~al.}(2013){Nikoli{\'c}}, {van de Ven}, {Heng}, {Kupko}, {Husemann}, {Raymond}, {Hughes}, \& {Falc{\'o}n-Barroso}}]{nikolic2013}
{Nikoli{\'c}}, S., {van de Ven}, G., {Heng}, K., {et~al.} 2013, Science, 340, 45, \dodoi{10.1126/science.1228297}

\bibitem[{{Ocker} {et~al.}(2021){Ocker}, {Cordes}, {Chatterjee}, \& {Dolch}}]{ocker_bowshocks}
{Ocker}, S.~K., {Cordes}, J.~M., {Chatterjee}, S., \& {Dolch}, T. 2021, ApJ, 922, 233, \dodoi{10.3847/1538-4357/ac2b28}

\bibitem[{{Orusa} {et~al.}(2021){Orusa}, {Manconi}, {Donato}, \& {Di Mauro}}]{orusa2021}
{Orusa}, L., {Manconi}, S., {Donato}, F., \& {Di Mauro}, M. 2021, JCAP, 2021, 014, \dodoi{10.1088/1475-7516/2021/12/014}

\bibitem[{{Petrov} {et~al.}(2020){Petrov}, {Bykov}, \& {Osipov}}]{petrov2020}
{Petrov}, A.~E., {Bykov}, A.~M., \& {Osipov}, S.~M. 2020, in Journal of Physics Conference Series, Vol. 1697, Journal of Physics Conference Series (IOP), 012002, \dodoi{10.1088/1742-6596/1697/1/012002}

\bibitem[{{Rangelov} {et~al.}(2017){Rangelov}, {Pavlov}, {Kargaltsev}, {Reisenegger}, {Guillot}, {van Kerkwijk}, \& {Reyes}}]{2017ApJ...835..264R}
{Rangelov}, B., {Pavlov}, G.~G., {Kargaltsev}, O., {et~al.} 2017, ApJ, 835, 264, \dodoi{10.3847/1538-4357/835/2/264}

\bibitem[{{Raymond} {et~al.}(2023){Raymond}, {Ghavamian}, {Bohdan}, {Ryu}, {Niemiec}, {Sironi}, {Tran}, {Amato}, {Hoshino}, {Pohl}, {Amano}, \& {Fiuza}}]{raymond2023}
{Raymond}, J.~C., {Ghavamian}, P., {Bohdan}, A., {et~al.} 2023, ApJ, 949, 50, \dodoi{10.3847/1538-4357/acc528}

\bibitem[{{Romani} {et~al.}(2010){Romani}, {Shaw}, {Camilo}, {Cotter}, \& {Sivakoff}}]{romani2010}
{Romani}, R.~W., {Shaw}, M.~S., {Camilo}, F., {Cotter}, G., \& {Sivakoff}, G.~R. 2010, ApJ, 724, 908, \dodoi{10.1088/0004-637X/724/2/908}

\bibitem[{{Romani} {et~al.}(2017){Romani}, {Slane}, \& {Green}}]{romani2017}
{Romani}, R.~W., {Slane}, P., \& {Green}, A.~W. 2017, ApJ, 851, 61, \dodoi{10.3847/1538-4357/aa9890}

\bibitem[{{Romani} {et~al.}(2022){Romani}, {Deller}, {Guillemot}, {Ding}, {de Vries}, {Parker}, {Zavala}, {Chalumeau}, \& {Cognard}}]{romani2022}
{Romani}, R.~W., {Deller}, A., {Guillemot}, L., {et~al.} 2022, ApJ, 930, 101, \dodoi{10.3847/1538-4357/ac6263}

\bibitem[{{Thaler} {et~al.}(2023){Thaler}, {Kissmann}, \& {Reimer}}]{thaler2023}
{Thaler}, J., {Kissmann}, R., \& {Reimer}, O. 2023, Astroparticle Physics, 144, 102776, \dodoi{10.1016/j.astropartphys.2022.102776}

\bibitem[{{van Adelsberg} {et~al.}(2008){van Adelsberg}, {Heng}, {McCray}, \& {Raymond}}]{vanAdelsberg2008}
{van Adelsberg}, M., {Heng}, K., {McCray}, R., \& {Raymond}, J.~C. 2008, ApJ, 689, 1089, \dodoi{10.1086/592680}

\bibitem[{{van Dokkum}(2001)}]{astroscrappy}
{van Dokkum}, P.~G. 2001, PASP, 113, 1420, \dodoi{10.1086/323894}

\bibitem[{{van Kerkwijk} \& {Kulkarni}(2001)}]{vanKerwijk2001}
{van Kerkwijk}, M.~H., \& {Kulkarni}, S.~R. 2001, A\&A, 380, 221, \dodoi{10.1051/0004-6361:20011386}

\bibitem[{{Verbunt} {et~al.}(2017){Verbunt}, {Igoshev}, \& {Cator}}]{verbunt2017}
{Verbunt}, F., {Igoshev}, A., \& {Cator}, E. 2017, A\&A, 608, A57, \dodoi{10.1051/0004-6361/201731518}

\bibitem[{{Vigelius} {et~al.}(2007){Vigelius}, {Melatos}, {Chatterjee}, {Gaensler}, \& {Ghavamian}}]{2007MNRAS.374..793V}
{Vigelius}, M., {Melatos}, A., {Chatterjee}, S., {Gaensler}, B.~M., \& {Ghavamian}, P. 2007, MNRAS, 374, 793, \dodoi{10.1111/j.1365-2966.2006.11193.x}

\bibitem[{{Wilkin}(1996)}]{1996ApJ...459L..31W}
{Wilkin}, F.~P. 1996, ApJL, 459, L31, \dodoi{10.1086/309939}

\bibitem[{{Wilkin}(2000)}]{wilkin2000}
---. 2000, ApJ, 532, 400, \dodoi{10.1086/308576}

\end{thebibliography}

\end{document}